\documentclass[aps,pre,epsf,twocolumn,amssymb,showpacs]{revtex4}
\usepackage{amsfonts}
\usepackage{graphicx}
\usepackage{graphics}
\usepackage{eepic,epsfig}
\usepackage{dcolumn}
\usepackage{bm}

\newcommand{\nn}{\nonumber}
\begin{document}
\title{Localization-delocalization transition in the quasi-one-dimensional
ladder chain with correlated disorder}
\author{Tigran Sedrakyan}
\email{tsedraky@ictp.trieste.it}
\author{Alexander Ossipov}
\affiliation{The Abdus Salam International Centre for
 Theoretical Physics,
Strada Costiera 11, I-34014 Trieste, Italy }
\date{\today}
\begin{abstract}
The generalization of the dimer model on a two-leg ladder is defined
and investigated both, analytically and numerically. For the closed
system we calculate the Landauer resistance analytically and found
the presence of the point of delocalization  at the band center which is
confirmed by the numerical calculations of the Lyapunov exponent. We calculate
also analytically  the localization length index and present the
numerical investigations of the density of states (DOS). For the open
counterpart of this model  the distribution of the Wigner delay times is
calculated numerically. It is shown how the localization-delocalization
transition manifest itself in the behavior of the distribution.

\end{abstract}
\maketitle

\section{\label{intro} Introduction}
One of the widely accepted results of the disordered systems is
that in one-dimensional spaces all the electronic states in
independent site-energy random (diagonal disorder) Anderson model
with non-random hopping amplitudes are exponentially localized
\cite{And58}. The diffusion or long-range transport of the initially localized
 particle in such systems is absent regardless the strength of the disorder.
The same result is  proved rigorously for quasi-one-dimensional
systems \cite{Efet}.

On the other hand, it is well established that the type of
disorder can have a strong effect on localization properties of
disordered systems. For instance, in one-dimensional systems with
off-diagonal disorder (random hopping models) delocalization takes
place at the band center \cite{Dyson}. In quasi-one-dimensional
systems with off-diagonal disorder the picture is more
complicated. It appeared \cite{Mud1, Mud2}, that in a weakly
disordered $N$-leg quasi-one-dimensional disordered tight-binding
hopping
 model there is a  delocalization transition at the critical energy
  $E_{crit}=0$  if  and only if $N$ is odd   (including $N=1$ case).
 The wave-functions remain localized for even $N$. The localization properties
of two- and three-channel tight-binding  Anderson model with nearest-neighbor
inter-chain hopping was studied in \cite{Hein}.

Another possibility  to have localization-delocalization transition in one
dimension is the correlated disorder, when the random variables are
 correlated at short or long distances.
There is intensive current interest both theoretical (see
\cite{SS3,DWP,7SS,SS6,SS9,DGK1,IK,ML,VP,P-D,SS1,SS,TS,DZL,DL,IKT95,IKT96} and
 Refs. therein) and experimental \cite{SS21,Stoeck} concerning analyze of
conditions, under which delocalized states can appear. This kind
of models with correlated disorder give explanation of high
conductivity of polymers such as doped poly-aniline and also
describe the transport properties of random  semiconductor
superlattices  \cite{P-D, SS, SS21}. Some transport properties of
many-mode waveguides with rough surfaces were studied in
\cite{MI} and the possibility of perfect transmission due to some specific
long-range correlations in the surface profiles was investigated. The scaling
properties of the models with correlated disorder were studied in
\cite{DZL, DL,IKT96}. In \cite{DZL} it was shown, that universality of
one parameter scaling breaks down in presence of long range
correlated disorder, for instance in the periodic-on-average
systems. It is necessary to point out, that the correlated
disordered systems do not belong to Zirnbauer's \cite{Zirn} and
Caselle's \cite{Cas} classification of fully disordered systems on
the basis of Cartan's classification of symmetric spaces. That
happens because, due to correlations, the Brownian motion of the
effective degrees of freedom carries out on the restricted part of
the coset space. This restriction is general characteristic of all
correlated disordered systems.

So far the discussion of correlated disorder was applied only to strictly
one-dimensional systems. In order to analyze the existence of
localization-delocalization transition in higher dimensional lattices with
correlated disorder,
it is natural to make the first step in this direction --- to
study the quasi-one-dimensional systems. To investigate the
existence of critical point and address the problem of
delocalization transition in this case analytically is an
important task. The exact calculation of Landauer localization
length and critical indices will allow us to understand whether
numerical simulation of Lyapunov exponent shows truly critical
point (where the localization length is infinite) or an extended
state (where the localization length is finite, but exceeds the
system size).

In this paper we will consider the extension of the random dimer
model (RDM) \cite{SS3,DWP,7SS,SS6,SS9,DGK1,IK,TS,SS21} to
quasi-one-dimensional case and show the existence of
delocalization transition for particular realization of
correlation. Namely, we consider a two-leg ladder chain (see
Fig.1) and a tight-binding model of random binary alloy with
on-site potential fields taking values ${\epsilon_a, \epsilon_b}$ from the
group $\Bbb{Z}_2$,  randomly assigned  to the lattice sites with probabilities
$p$ and $(1-p)$, correspondingly.
Here we will focus on a particular case, when the same site
potentials appear always by quartets disposed on four nearest
neighbor sites creating squares, as it is marked on the Fig.1.
This is most natural generalization of the RDM in case of ladder
models, and one can call it random ladder dimer model (RLDM).


\begin{figure}
\includegraphics{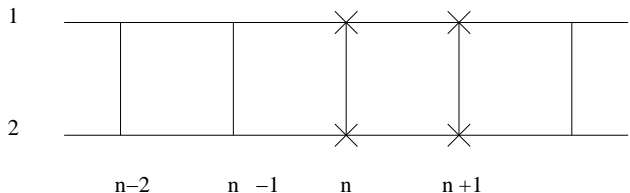}
\caption{Two leg ladder}\label{Fig1}
\end{figure}

The  Hamiltonian of the RLDM is given by
\begin{eqnarray}
\label{H}
H=\sum_{n=1; i=1,2}^{N}\{ t_h (c_{n,i}^+ c_{n+1,i}&+&c_{n+1,i}^+c_{n,i})\nn\\
+t_v c_{n,i}^+c_{n,i+1}
&+&\epsilon_{n,i}c_{n,i}^+c_{n,i}\}.
\end{eqnarray}
Here $t_v$ and $t_h$ are constant vertical and horizontal hopping
parameters correspondingly, $\epsilon_{n,i}$ is the random
external potential at site $n$ of the $i$-th chain. $N$ denotes
the number of atoms in  each  chain. We impose a constraint on the
randomness  by demanding that four $\epsilon_{n,i} \;  (n=2k-1,
2k; \; i=1,2)$ around the plaquette (see Fig.1) always have the
same, but randomly chosen value from the set $\left\lbrace \pm
m\right\rbrace$
\begin{eqnarray}
\label{eps}
\epsilon_{2 k-1,1}=\epsilon_{2 k-1,2}=\epsilon_{2 k,1}=
\epsilon_{2 k,2} \in \left\lbrace
\pm m\right\rbrace, \nonumber \\
k=1,\dots ,{N \over 2}.
\end{eqnarray}

In section \ref{closed} we study the closed system described by
Hamiltonian (\ref{H}). We use the technique introduced in
\cite{P-D},  \cite{SS1}, \cite{SS} and further applied to the RDM
in \cite{TS} in order to investigate the RLDM. We calculate
analytically the dimensionless Landauer exponent of the model and
compare it with numerical simulations of the Lyapunov exponent
(Fig.2 and Fig.3), calculated iteratively in standard way
\cite{CPV}, \cite{MacK1}. The exact calculation of the Landauer
resistance shows the existence of one real critical point with
critical energy $E_{crit}=0$ in the case of
$m = t_v/t_h$. Otherwise all
states are localized.

In the case of ${m=t_v/t_h < 1}$ we analytically analyze the
divergence of the inverse of Landauer exponent and find, that
there is delocalization with the critical index $\nu=2$. On Fig.2
and Fig.3 we see very good coincidence of the half of Landauer
exponent with the Lyapunov one around the critical point.

The critical behavior is essentially different when
${m=t_v/t_h=1}$. In this case also the localization length
diverges at the point $E_{crit}=0$, but with index $\nu=1/2$. The
comparison of Landauer and Lyapunov exponents for this case is
presented on Fig.4.

 The numerical analyze of the density of states
(DOS) shows anomalous behavior at the band center in the case,
when ${t_v=t_h}$ (see Fig.5) and regular behavior for ${t_v<t_h}$
(see Fig.6).

Section \ref{open} is devoted to the open counterpart of the
system described by (\ref{H}). There we show  that far from the
delocalization point  the distributions  of the Wigner delay times
are very similar to those found for the random models with
uncorrelated disorder (Fig.~\ref{fig1} and Fig.~\ref{fig2}). At
the critical point the behavior of the distribution depends on
$m$. Namely, for $m=1$ the distribution of the re-scaled Wigner
delay times tends to the $\delta$-function (Fig.~\ref{fig3}),
corresponding  to the deterministic, ballistic propagation. While
for $m<1$ this distribution is not deterministic, but still
bounded (Fig.~\ref{fig4}),  indicating the ballistic propagation
with velocities distributed in some interval.

\section{\label{closed} Landauer resistance, critical exponents and density of states}

For one particle eigenstates $\psi_n(E)$ of the eigenenergy $E$
the Schr\"odinger equation of the Hamiltonian (\ref{H}) has the following recursion form:
\begin{eqnarray}
\label{shr}
t_h (\psi_{n+1,i}+\psi_{n-1,i})+t_v\psi_{n,i+1}=(E-\epsilon_{n,i})\psi_{n,i}.
\end{eqnarray}
Now we can rewrite (\ref{shr}) in the matrix form
\begin{eqnarray}
\label{T}
 \left(
\begin{array}{l}
\psi_{n+1,1} \\
\psi_{n+1,2}\\
\psi_{n,1} \\
\psi_{n,2}
\end{array}
\right) =
T_n \left(
\begin{array}{l}
\psi_{n,1} \\
\psi_{n,2} \\
\psi_{n-1,1} \\
\psi_{n-1,2}
\end{array}
\right),
\end{eqnarray}
by introducing $4\times 4$ transfer matrix $T_n$ as
\begin{eqnarray}
\label{T1}
T_n =  \left(
\begin{array}{cccc}
E-\epsilon_{n,1} & -t & -1 & 0  \\
-t &  E-\epsilon_{n,2} & 0 & -1 \\
1 & 0 & 0 & 0 \\
0 & 1 & 0 & 0
\end{array}
\right).
\end{eqnarray}
Here we  re-scaled the energies and the vertical hopping parameter
$t_v$ (introducing $t$ instead) by the horizontal hopping
parameter $t_h$. One can easily find out, that the $4\times 4$
transfer matrix $T_n$ possesses the condition
  $T_n^+ J T_n= J$ with
\begin{eqnarray}
\label{J}
J=  \left(
\begin{array}{cccc}
0 & 0 & -i & 0  \\
0 & 0 & 0 & -i \\
i & 0 & 0 & 0 \\
0 & i & 0 & 0
\end{array}
\right)
\end{eqnarray}
which means that $T_n$ is an element of the group $SU(2,2)$.
If we introduce a new matrix $M_n=\prod_{i=n}^{1}T_i$, then we can relate
the wave functions
$(\psi_{n+1,1},\psi_{n+1,2},\psi_{n,1},\psi_{n,2})$ with ones at sites 0
and 1 of both chains
\begin{eqnarray}
\label{prom}
\left(
\begin{array}{l}
\psi_{n+1,1} \\
\psi_{n+1,2}\\
\psi_{n,1} \\
\psi_{n,2}
\end{array}
\right) = M_n
\left(
\begin{array}{l}
\psi_{1,1} \\
\psi_{1,2}\\
\psi_{0,1} \\
\psi_{0,2}
\end{array}
\right),
\end{eqnarray}
and $M_N$ will denote the total transfer matrix of the system.
From the definition of $M_N$ it is obvious, that it has the same
property $M_N^+J M_N=J$. This condition is equivalent to the flux
conservation in the theory (i.e., Hermiticity of $H$).

In a physics of disorder particular interests has  the limit of
eigenvalues of the matrix $\ln(M_NM_N^+)^{1\over{2N}}$ when
$N\rightarrow\infty$. According to Oseledec theorem \cite{O} this
limit exists and the maximal eigenvalue is called
Lyapunov exponent.  Due to self averaging property it is determined
by the maximal eigenvalue of the matrix $M_N$
\begin{equation}
\label{Lyap}
\gamma_{Lyapunov}=\lim_{N \rightarrow
\infty}\langle\ln \mid \mid M_N\mid \mid^{1\over{N}}\rangle,
\end{equation}
where $||M_N||^2=\sum_{\alpha,\beta}(M_{N})_{\alpha}^{\beta}
(M_N^*)_ {\alpha}^{ \beta}=Tr(M_N M_N^+)$ and the average is taken over random
distribution of the external potential taking into account the
correlations (if they are present in the model under consideration).

It is hard to calculate the Lyapunov exponent exactly and
establish the presence of delocalization point. Another variable,
which shares the essential characteristics of the
localization-delocalization transition but can be calculated
exactly is the dimensionless Landauer resistance, which is defined
as  the ratio of squares of the modules of the reflection and
transmission amplitudes. According to arguments of Anderson et
al. \cite{ATAF} in one-dimensional systems the Landauer exponent (defined in Eq. (\ref{ld}))
is simply the double of Lyapunov one.  The proportionality of these two
exponents  was confirmed  around of the critical point in the paper
\cite{SSS}. Considering in most general terms one dimensional
models with transfer matrices belonging to $SU(1,1)$ authors of
\cite{SSS} have shown,  that in the lowest order of perturbation expansion
with some parameter  measuring  the distance from the critical point,
the Landauer exponent is twice the Lyapunov
one, just as predicted by \cite{ATAF}. This means, that critical indices of
both exponents are equal. In quasi-one dimensional models, which   have
transfer  matrix from $SU(n,n)$, again one can expect the coincidence of
critical indices.

The exponential increase of Landauer resistance with distance is determined by
the  Landauer exponent , which can be defined as
\begin{equation}
\label{ld}
\gamma(E)= \lim_{N \rightarrow \infty}\ln \langle\mid\mid M_N\mid
\mid \rangle^{1\over{N}}.
\end{equation}
In the articles \cite{P-D}-\cite{SS}  it was shown, that the Landauer
resistance (and
corresponding exponent) can be calculated exactly
just by reducing the direct product $\ M_N\otimes M_N^+$ of fundamental
representations of
transfer matrices $M$  of $SU(1,1)$ group
to the adjoint one. We apply this technique here for the group $SU(2,2)$.
In order to calculate this direct product exactly, we  use the known
representation of the permutation operator
 via generators $\tau^\mu$  $ (\mu =1,\dots,15)$ of the $sl(4) $ algebra
as $P={1\over 4} (\Bbb{I} + \tau^{\mu}\otimes\tau^{\mu})$. In the matrix
elements
this formula looks like
\begin{eqnarray}
\label{Perm}
\delta_{\alpha_1}^{\alpha_2}\delta_{\beta_1}^{\beta_2}={1\over 4}
[\delta_{\alpha_1}^{\beta_2}\delta_{\beta_1}^{\alpha_2}+
(\tau^\mu)_{\alpha_1}^{\beta_2}(\tau_\mu)_{\beta_1}^{\alpha_2}],
\end{eqnarray}
where  we suppose summation by the repeating indices $\mu$.
Among of generators $\tau^\mu$ there is one, which coincides with the
metric $J$
defined in (\ref{J}). We denote the corresponding index $\mu$ as $J$, namely
$\tau^J = J$.

\begin{figure}
\includegraphics{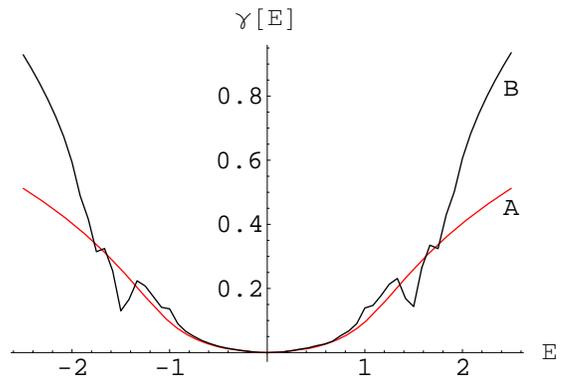}
 \caption{The Lyapunov (curve B) and the half of
Landauer (curve A) exponents versus energy in the whole interval \newline
[-2.5,2.5] of existing eigenstates in case of $m=t_v/t_h$=1/2. The
chains of the length N=40000 at 60 values of energies were
considered. \label{Fig2a}}
\end{figure}
\begin{figure}
\includegraphics{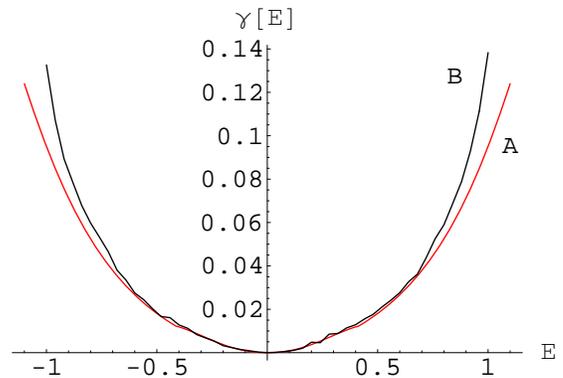}
 \caption{The Lyapunov (curve B) and the half of Landauer (curve A) exponents
in the vicinity of the critical point E=0 versus energy E in case of
$m=t_v/t_h=1/2$. The chains of the length N=40000 for 60 values of
energies in the interval [-1.1, 1.1] were considered
\label{Fig2b}}
\end{figure}

By simple multiplication of (\ref{Perm}) by $T_j$ and $T^+_j$ from
the left and right hand sides correspondingly one can express the
direct product of $T_j$ and $T^+_j$ via their adjoint
representation:
\begin{eqnarray}
\label{TT}
(T_j)^{\alpha}_{\alpha^{\prime}}(T_{j}^{+})^{\beta^{\prime}}_{\beta}=
{\frac{1}{4}}(J)^{\alpha}_{\beta}
(J)^{\beta^{\prime}}_{\alpha^{\prime}}+ {\frac{1 }{4}}(\tau^{\mu}
J)^{\beta^{\prime}}_{\alpha^{\prime}} \Lambda_{j}^{\mu\nu}(J
\tau^{\nu})^{\alpha}_{\beta}.
\end{eqnarray}

Here the adjoint representation $\Lambda_n$  of $T_n$ is defined
by
\begin{eqnarray}
\label{Lambda} \Lambda_n^{\mu\nu}={1\over 4}{\rm
Tr}(T_n\tau^{\mu}T^{+}_n\tau^{\nu})
\end{eqnarray}
and is an $15\times 15$ matrix depending on the parameters of the model at the
site $n$ of both chains.

Now, in order to calculate the average of Landauer exponent, we
use (\ref{TT}) and decompose the direct product of two transfer
matrices corresponding to the whole system of $2N$ sites. In the
model under consideration we attach the same site potential
$\epsilon$ to the quartet of sites (see Fig.1) and calculate the
average over random distribution of values $\epsilon_a$ and
$\epsilon_b$ with probabilities $p$ and $1-p$ respectively.
Therefore we should consider the $\Lambda^2$ as a constituent
transfer matrix and average it in a product

\begin{equation}
\langle M_NM_N^+ \rangle ={\frac{1 }{4}} J \otimes J +
{\frac{1}{4}} (\tau^{\mu} J)\otimes (J \tau^{\nu})
\left(\prod_{j=1}^{N/2} \langle\Lambda_{j}^2\rangle\right)^{\mu
\nu}. \label{Gamma}
\end{equation}

It is clear, that
\begin{eqnarray}
\label{L2} (\Lambda_{j}^2)^{\mu\nu}={\frac{1}{2}}\,{\rm
Tr}\left(T_j^2\tau^{\mu} [T_{j}^2]^{-1} \tau^{\nu}\right)
\end{eqnarray}
and the average is
\begin{eqnarray}
\label{L22}
\Lambda =\langle\Lambda_n^2\rangle =
p\Lambda_n^2(\epsilon_a)+(1-p)\Lambda_n^2(\epsilon_b),
\end{eqnarray}
where $\Lambda_j(\epsilon_a)$ and $\Lambda_j(\epsilon_b)$ are calculated for the site
potentials $\epsilon_a$ and $\epsilon_b$ respectively. The matrix elements of
$\Lambda$ can be found easily from (\ref{L2}) and (\ref{L22}).
Hereafter we take $\epsilon_a=-m$, $\epsilon_b=+m$ without lose of generality
and  $p=1/2$ for simplicity.

From the formula (\ref{Gamma}) one can easily obtain
\begin{eqnarray}
\label{MMM} \langle\mid\mid M\mid\mid^2\rangle= 4
(\Lambda^{N/2})^{J J}\simeq\lambda_{max}^{N/2},
\end{eqnarray}
where the biggest eigenvalue $\lambda_{max}$ defines the asymptotic
behavior of Landauer resistance with the distance. In (\ref{MMM})
superscript  $J J$ means $JJ$ element of the matrix $\Lambda^{N/2}$
(see notion after formula (\ref{Perm})).

According to (\ref{ld}) the Landauer exponent $\gamma(E)$
connected with the maximal eigenvalue as
\begin{equation}\label{le}
    \gamma(E)= \frac{1}{4} \log[\lambda_{max}]
\end{equation}
and defines the inverse of Landauer localization length $\xi(E)= 1/
\gamma(E)$.

\begin{figure}
\includegraphics{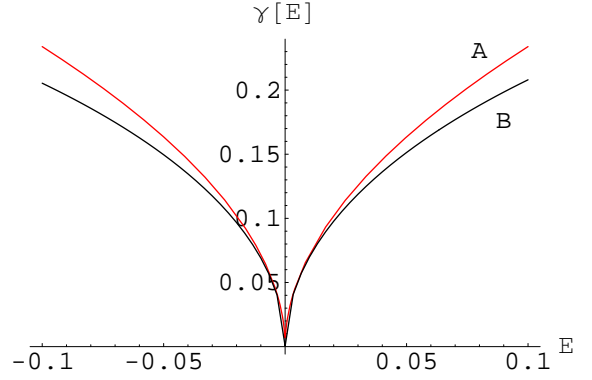}
\caption{The Lyapunov (curve B) and Landauer (curve A) exponents in
close vicinity of the critical point E=0 versus energy E in case
of $m=t_v/t_h=1$. The chains of the length N=40000 for 60 values
of energies in the interval [-0.1, 0.1] were
considered}\label{Fig3}
\end{figure}


The delocalized states correspond to the critical points where
$\xi(E_c)=\infty$, or $\lambda_{max}=1$,  and can be found by the condition
$det\mid {\Bbb I}-\Lambda \mid =0\nn$. In our case this characteristic
equation is given as
\begin{equation}
\label{character}
m^4[(E-t)^2-m^2][(E+t)^2-m^2][m^4-E^2(-1+m^2)]t^2=0.
\end{equation}
From this equation one can find following candidates for critical
points
\begin{eqnarray}
\label{Ec1}
E_{1c}&=&t \pm m,\\
\label{Ec2}
\quad E_{2c}&=&-t \pm m,\\
\label{Ec3}
 E_{3c}&=&m^4/(m^2-1).
\end{eqnarray}

\begin{figure}
\includegraphics{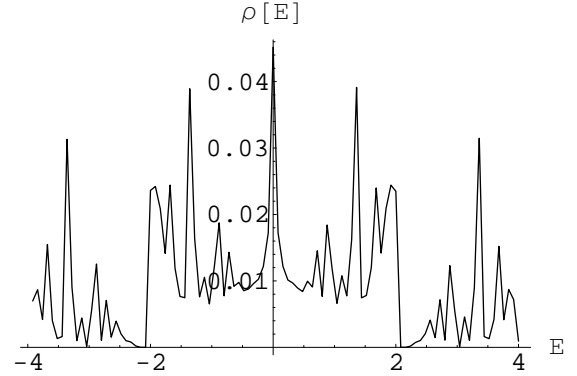}
\caption{The density of states dependence on energy for the
boundary case $m=\epsilon_a-\epsilon_b=2$ calculated numerically
for the ensemble of 4000 coupled chains with $N=50$ sites in each
chain.} \label{Fig4}
\end{figure}

But since the Landauer exponent defined by the maximum eigenvalue
of $\Lambda$ one need to find out whether this points belong to
maximum eigenvalue or not. The full characteristic equation
$det\mid {\lambda \Bbb I}-\Lambda \mid =0$ for eigenvalues
$\lambda$ factorizes into the product of three polynomials
$U_1(\lambda, E,t,m) U_2(\lambda, E,t,m) \left ( U_3(\lambda, E,t,m)\right )^2$ and
gives us
\begin{eqnarray}
\label{U1}
U_1(\lambda,E,t,m)=(1-2m^2)^2+\left[(2m^2-1)(3-\right.\nn\\
\left. 2m^2+(2E+E^2-m^2-2(1+E)t+t^2)(E^2-\right.\nn\\
\left. m^2-2E(1+t)+t(2+t)))\right] \lambda +\left[3+E^4+\right.\nn\\
\left. m^4-4E^3t-4t^2+t^4-4Et(3m^2+t^2-2)+\right.\nn\\
\left. m^2(6t^2-4)+E^2(6m^2+6t^2-4)\right]\lambda ^2 -\lambda ^3=0\nn\\
\\
\label{U2}
U_2(\lambda, E,t,m)=(1-2m^2)^2+\left[(2m^2-1)(3-\right. \nn\\
\left. 2m^2+(2E+E^2-m^2+2(1+E)t+t^2)(E^2-\right. \nn\\
\left. m^2-2E(1-t)-t(2-t)))\right] \lambda +\left[3+E^4+\right.\nn\\
\left. m^4+4E^3t-4t^2+t^4+4Et(3m^2+t^2-2)+\right. \nn\\
\left. m^2(6t^2-4)+E^2(6m^2+6t^2-4)\right]\lambda ^2 -\lambda ^3 = 0,\nn\\
\\
\label{U3}
U_3(\lambda, E,t,m)=-(1-2m^2)^2-(2m^2-1)\left [E^4+m^4+\right. \nn\\
\left (t^2-2)^2-2E^2(2+m^2+t^2)+ m^2(6t^2-4)\right ]\lambda+ \nn\\
2\left [6m^2-3-3m^4+m^6+E^4(m^2-1)-2(m^4+ \right. \nn\\
\left. 3m^2-2)t^2+(m^2-1)t^4-2E^2(m^4-2+3t^2-\right. \nn\\
\left. 3m^2(t^2-1))\right ]\lambda ^2 +\left [E^4+m^4+(t^2-2)^2-\right. \nn\\
\left. 2m^2(2+t^2)+2E^2(3m^2-t^2-2)\right ]\lambda
^3-\lambda^4=0.\nn\\
\end{eqnarray}
Easy to see, that $U_2(\lambda, E,t,m)=U_1(\lambda, E,-t,m)$
The critical points $E_{1c},\; E_{2c}$ and $E_{3c}$ are zeros of
$U_1(\lambda, E,t,m),\;U_2(\lambda, E,t,m)$ and
$U_3(\lambda, E,t,m)$ respectively.

\begin{figure}
\includegraphics{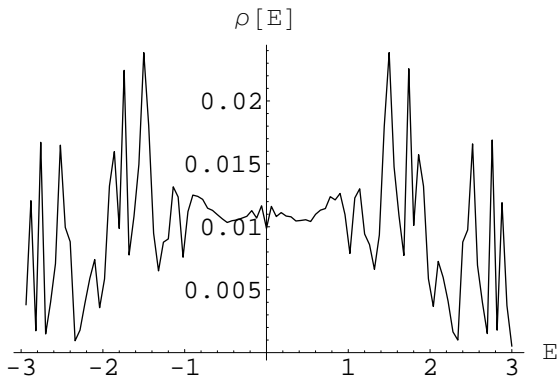}
\caption{The density of states dependence on energy for the case
$m=(\epsilon_a-\epsilon_b)/2=0.5$ calculated numerically for the
ensemble of 2000 coupled chains with $N=50$ sites in each
chain.}\label{Fig5}
\end{figure}

Cubic and quartic equations (\ref{U1}-\ref{U3}) can be solved
analytically. It appears, that the roots of quartic equation
(\ref{U3}) are either less of one or complex valued, therefore
they are out of interest. Roots of cubic equation $U_2(\lambda,
E,t,m)=0$ can be obtained from $U_1(\lambda, E,t,m)=0$ by simple
change $t \rightarrow-t$. Both of them behave as roots of the
corresponding equation in  one-channel RDM (see formula (14) and
Fig.1 in \cite{TS} ) but are shifted on $\pm t$ respectively.
Therefore, in general situation ($t\neq m$), the roots of 
one characteristic polynomial
are in a shadow of the brunch of the other polynomial and the maximum eigenvalue is always more
than one. But if one will fine tune $m=t$ , then from the
expressions (\ref{Ec1}-\ref{Ec2}) of $E_{1c}$ and $E_{2c}$ it
follows, that the point $E_c=0$ is common critical point and,
therefore, it is a critical point of maximum eigenvalue
$\lambda_{max}$. In the limiting case $t \rightarrow 0$, when the
transverse hopping is absent and we have two decoupled dimer
chains, there is no any shift of two curves, they just
coincide. Then we obtain the same delocalization points $\pm m$,
defined by formulas (\ref{Ec1}-\ref{Ec2}), as in one-channel RDM.
We find also, as in one-channel RDM \cite{SS3,DWP,7SS,IK,TS},
the presence of critical point only for case of $m\leq 1$, when
each of branches can have critical points (\ref{Ec1}) and
(\ref{Ec2}).

The presence of delocalization point $E=0$ can be proved
rigorously analyzing the Lyapunov exponent. We have two transfer
matrices (depending on $\pm m$) randomly distributed in a product
along a chain.
The maximal eigenvalue of this product after statistical averaging  just
 defines the Lyapunov exponent (\ref{Lyap}) of the model.
 If there is a particular value
of $E$, where this two transfer matrices commute (this means
all matrices in the product commute) then the eigenvalues of
the product are given by  the product of the corresponding eigenvalues of
 the constituent random matrices. If, at the same time,  the eigenvalues
of both transfer matrices at this point are laying on the unit circle
 (modules of eigenvalues are equal to one), we have a  point of delocalization.
In order to see that, let us diagonalize the transfer matrix $M_N$
by the matrix $S$, which is also diagonalizing matrix of the individual
$T_i$ 's along a chain.  One  obtains $\gamma_{Lyapunov}=\lim_{N \rightarrow
\infty}{1\over{2 N}}\ln Tr(S M_d S^{-1} (S^{-1})^+ M_d^+ S^+)$, where
the eigenvalues in $M_d$,  as a product of individual eigenvalues
of $T_i$,  also have module equal to one. Therefore, the argument
of logarithm is bounded from above by some constant independent of $N$
and the limit $N\rightarrow\infty$ is zero.
It remains now to check, that indeed the squares of transfer matrices
(\ref{T1}) for $\epsilon_{n,i}=\pm m$ are commuting at $E=0$ and for $t=m$.
For $t \leq 1$ they have two eigenvalues equal to $-1$ and two complex
conjugate eigenvalues on the unit circle ($e^{\pm i \alpha}$).
This prove the  presence of the delocalization point.

In Fig.2. we present the graphic of the half of Landauer exponent $\gamma(E)$
versus $E$ (curve A), defined by the maximum solution of cubic
equations (\ref{U1}-\ref{U3}) for the generic value $m=t_v/t_h=1/2$. We
present here also the Lyapunov exponent $\bar{\gamma}$
(curve B) obtained by the numerical simulations of the chains of the
length $N=40000$ at 60 values of energies in the whole energy interval
$[-2.5,2.5]$. The  same exponents in the vicinity of critical point
($E \in [-1.1,1.1]$) are presented on the Fig.3 in order to demonstrate
the coincidence of critical behaviors, as it expected after the articles
\cite{ATAF} and \cite{SSS}.  The critical index $\nu$ is 2 in this case.

In Fig.4 we present the  Landauer and the Lyapunov exponents versus $E$
for the boundary case $m=t_v/t_h=1$.  We present results for  close vicinity
of $E=0$  and one can see exactly the same critical behavior of the both exponents.
It is necessary to mention, that in contrast to the generic case $m \neq 1$ here the
multiplication factor 1/2 is absent.

For $m>1$ the delocalization point $E_c=0$ disappears.

The critical index $\nu$ of the Landauer exponent one can obtain by
expanding  the solutions of the equations
(\ref{U1}-\ref{U2}) near the critical point $E_c=0$
\begin{eqnarray}
\label{nu} \gamma(E)=\left\{\begin{array}{l} {1\over 6}E^2+{1\over
9}E^3+O[E]^4, \quad \text{if} \quad m=t={1\over 2}\\
\\ \sqrt{E\over 2}-{1\over 8}E+O[E]^{3/2}, \quad \text{if} \quad
m=t=1.
\end{array}
\right.
\end{eqnarray}
As we see the critical indices are different for generic $m=1/2,
\; \nu=2$ and boundary $m=1,  \; \nu=1/2$ cases.  Fig.3 and Fig.4
demonstrate that difference. The similar situation was in the
one-channel RDM \cite{TS}. The index $\nu=1/2$ is equal to the
corresponding index in one-channel RDM in the same boundary case
$m=1$, when one is approaching to the critical points $E_c=\pm 1$ from
the outside of the region $[-1,1]$ (see \cite{TS} ).  After the discussion presented
above this fact is natural to expect.

In order to fill up the exact results from above, we have analyzed
the density of states in the model by the numerical
diagonalization of the Hamiltonian (\ref{H}).

In Fig.5 the DOS versus energy for the boundary case
$m=\epsilon_a-\epsilon_b=2$ is presented. Here we calculated
the DOS for a system with $N=50$ sites in each chain. The result
comprise the average over 4000 realizations of the disorder. Here
one can clearly see the anomaly of DOS at band center $E=0$
(in accordance with \cite{KW}),
corresponding to delocalization transition. A similar anomaly is
present in RDM at the both critical energies $\epsilon_a,
\epsilon_b$ in the same case when $\epsilon_a-\epsilon_b=2$.

In the Fig.6 we present the normalized DOS for the generic case
$m=0.5$, calculated numerically for the ensemble of 2000 coupled
chains of length $N=50$ each. This case corresponds to a different
type of critical behavior with respect to the previous one and the
DOS is finite in whole energy interval.

\section{\label{open} Wigner delay time}

In this section we consider the properties of the open counterpart of the
model studied in section \ref{closed} . Namely we investigate the statistical
distributions of the Wigner delay times for various regimes. The Wigner delay
time is defined as the derivative of the total phase of the scattering matrix
with respect to the energy $\tau=-i\frac{\partial {\rm ln \; det}
  S(E)}{\partial  E}$. This quantity was intensively studied in recent years:
for chaotic \cite{FS97}, localized \cite{OKG00,JVK89, F03}, diffusive
\cite{OKG03, F03} systems as well as for systems at criticality
\cite{SOKG00, F03}.

\begin{figure}[t]
\epsfig{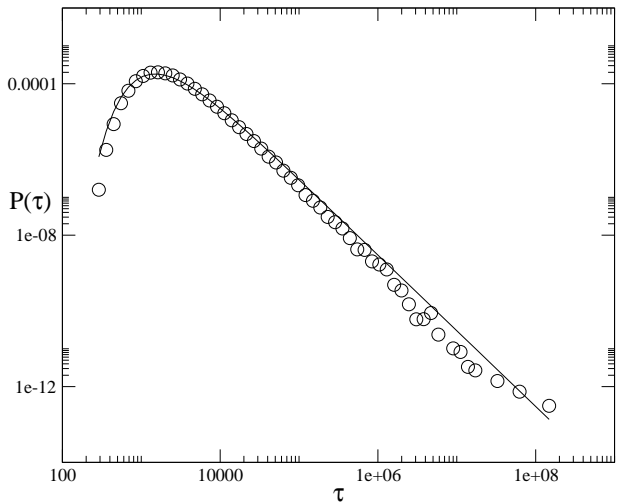}
\caption{\label{fig1} Distribution of the Wigner delay times $\tau$ for
  $E=0.3$, $t=1$, $m=0.1$ and $N=10^4$ (circles). The line represents the best
  fit by formula (\ref{ander}) with the fit parameter $a$.}
\end{figure}

In order to study an open system we attach a perfect lead to the semi-infinite
disordered sample. Thus $\epsilon_{n,i}=0$, $t=1$ for $n\leq 0$ and they take
the same values as in the previous section for $n>0$. Let us first assume more
general situation of $M$ channels and derive the expression for the Wigner
delay time.

The wavefunction in the lead is a superposition of the plane waves in the
longitudinal direction and the standing waves in the transverse direction:
\begin{equation}
\psi_{n,i}=\sum_{j=1}^M\left\{\frac{A_j}{\sqrt{\sin k_j}}e^{ik_j n}\phi_i^j+
\frac{B_j}{\sqrt{\sin k_j}}e^{-ik_j n}\phi_i^j\right\},
\end{equation}
where $\phi_i^j=\frac{1}{\sqrt{M}}\sin\left(\frac{\pi j}{M+1}i\right)$ is the
transverse eigenfunction and $k_j$ is the wave-vectors of $j$th mode determined
by the equation $E=2t_h\cos k_j+2t_v\cos\left(\frac{\pi j}{M+1}\right)$. In
order to decouple different modes it is useful to introduce new variables:
\begin{equation}
\vec{W}_n=U\vec{\psi}_n,
\end{equation}
where $\vec{W}_n=(W_{n,1},\dots,W_{n,M})^T, \; \vec{\psi}_n=(\psi_{n,1},\dots,
 \psi_{n,M})^T$ and $U$ is $M\times M$ matrix with elements $U_{l,i}=
 \phi^{l\ast}_i$. In new variables each component of the wavefunction in the
 lead contains contribution of one particular mode only:
\begin{equation}
W_{n,l}=\frac{A_l}{\sqrt{\sin k_l}}e^{ik_l n}+
\frac{B_l}{\sqrt{\sin k_l}}e^{-ik_l n}
\end{equation}
In order to rewrite (\ref{T}) in terms of the new variables we notice that the
generalization of the transfer matrix (\ref{T1}) for $M$ channel case is
$2M\times 2M$ block-matrix:
\begin{eqnarray}
T_n= \left(
\begin{array}{cc}
C_n &  -\Bbb I\\
\Bbb I & \Bbb O
\end{array}
\right),
\end{eqnarray}
where $C_n$ is a three-diagonal matrix generalizing the left upper block of the
transfer matrix (\ref{T1}). Now the analog of  (\ref{T}) in new variables
takes the similar form:
\begin{eqnarray}
\left(
\begin{array}{l}
\vec{W}_{n+1} \\
\vec{W}_{n}
\end{array}
\right) =
\tilde{T}_n \left(
\begin{array}{l}
\vec{W}_{n} \\
\vec{W}_{n-1}
\end{array}
\right),
\end{eqnarray}
where the new transfer matrix $\tilde{T}_n$ is giving by
\begin{eqnarray}
\label{newT}
\tilde{T}_n= \left(
\begin{array}{cc}
\tilde{C}_n &  -\Bbb I\\
\Bbb I & \Bbb O
\end{array}
\right), \;\;\;\; \tilde{C}_n\equiv UC_nU^{-1}
\end{eqnarray}
Let us now consider the finite sample of length $N$ and impose  Dirichlet
boundary condition at $n=N+1$. Then the analog of  equation (\ref{prom}) reads
\begin{eqnarray}
\label{forr}
\left(
 \begin{array}{l}
\vec{\Bbb O}\\
\vec{W}_N
\end{array}
\right)=\left(
\begin{array}{cc}
P^{11}_N & P^{12}_N\\
P^{21}_N & P^{22}_N
\end{array}
\right)\left(
\begin{array}{l}
\frac{1}{\sqrt{\sin K}}\left(\vec{A}_N+\vec{B}_N\right)\\
\frac{1}{\sqrt{\sin K}}\left(e^{-iK}\vec{A}_N+e^{iK}\vec{B}_N\right)
\end{array}
\right),
\end{eqnarray}
where $P_N$ is the total transfer matrix $P_N=\prod_{n=1}^{N}\tilde{T}_n$ and
matrix $K$ is diagonal $K={\rm diag}(k_1,\dots,k_M)$.

\begin{figure}[t]
\epsfig{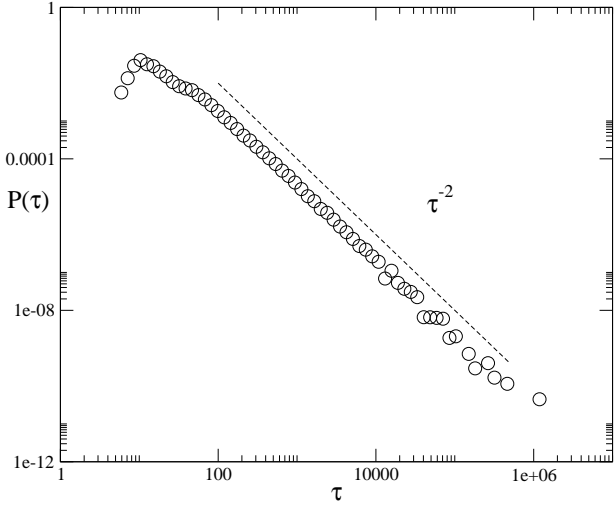}
\caption{\label{fig2} Distribution of the Wigner delay times $\tau$ for
  $E=0.3$, $m=t=1$,  and $N=10^4$ (circles). The dashed line represents
  $1/\tau^2$ behavior.}
\end{figure}

The scattering matrix for our geometry consists of only one non-trivial block
--- the reflection matrix $R_N$, which relates the incoming wave amplitudes
$\vec{A}_N$ to the outgoing wave amplitudes $\vec{B}_N$:
\begin{equation}
\vec{B}_N=R_N\vec{A}_N
\end{equation}
Using this definition and relation (\ref{forr}) one can easily calculate the
reflection matrix:
\begin{eqnarray}
R_N&=&-\left[\Bbb I+\sqrt{\sin K}F_N\left(\sqrt{\sin K}\right)^{-1}e^{iK}
\right]^{-1}\times\nonumber\\
&&\left[\Bbb I+\sqrt{\sin K}F_N\left(\sqrt{\sin K}\right)^{-1}
 e^{-iK}\right]\nonumber\\
F_N&\equiv& \left(P^{11}_N\right)^{-1}P^{12}_N
\end{eqnarray}
Taking the derivative with respect to energy from this expression one can find
the formula for the Wigner delay time:
\begin{equation}
\label{tau_N}
\tau_N=2{\rm Im \,Tr}\left[\Bbb I +F_Ne^{-iK}\right]^{-1}
\left[\frac{\partial F_N}{\partial E}e^{-ik}-iF_Ne^{-iK}\frac{\partial
    K}{\partial E}\right]
\end{equation}
The matrix $F_N$ and its derivative with respect to energy can be calculated
recursively. To this end we increase the length of the sample by one and see
how the total transfer matrix changes. Then from the definition of $P_N$
and Eq.~(\ref{newT}) one can derive the following recursion relations:
\begin{eqnarray}
\label{rec_rel}
F_{N+1}&=&-(\tilde{C}_N+F_N)^{-1}\nonumber\\
\frac{\partial F_{N+1}}{\partial E}&=&F_{N+1}\left[\Bbb I +\frac{\partial
    F_N}{\partial E}\right]F_{N+1}
\end{eqnarray}
Starting from the initial values $F_1=-(\tilde{C}_1)^{-1}$, $ \partial
F_1/\partial E=F_1^2$ and iterating  relations (\ref{rec_rel}) one can
calculate $F_N$ and $\partial F_N/\partial E$ for any system size $N$.
Inserting these values into Eq.~(\ref{tau_N}) we obtain the Wigner delay time
$\tau_N$.

\begin{figure}[t]
\epsfig{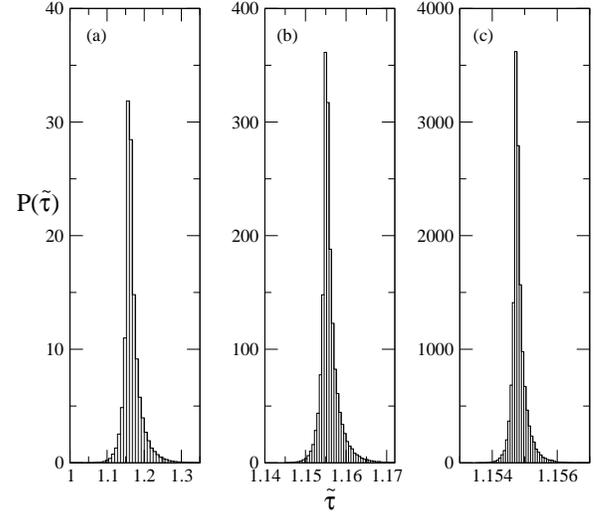}
\caption{\label{fig3} Distribution of the rescaled Wigner delay times
  $\tilde{\tau}\equiv\tau/N$ for $E=0$, $m=t=1$, and various system sizes:
  (a) $N=100$, (b) $N=1000$, (c) $N=10000$.}
\end{figure}

Below we present the results of numerical calculations of the distribution of
$\tau_N$ for the random ladder dimer model. In these particular case the
matrix $\tilde{C}_n$ becomes diagonal:
\begin{eqnarray}
\tilde{C}_n=\left(
\begin{array}{cc}
E-t-\epsilon_n &  0\\
0 & E+t-\epsilon_n
\end{array}
\right),
\end{eqnarray}
therefore the recursion relations (\ref{rec_rel}) are reduced to two decoupled
equations. That makes the numerical simulations especially efficient: in all
calculations we used $10^5$ realizations of the random potentials and the
system size up to $N=10^4$.

Let us now using the knowledge gained in the previous section  consider
 different regimes. We start with  the localized regime. In Fig.\ref{fig1} the
 distribution of the Wigner delay times is presented for $t=1$, $m=0.1$ and
 $E=0.3$. The distribution is described very nice by the formula:
\begin{equation}
\label{ander}
{\cal P}(\tau)=\frac{a}{\tau^2}e^{-a/\tau},
\end{equation}
which is derived for the distribution of $\tau$ for a one-dimensional Anderson
model in the weak disorder limit \cite{OKG00, JVK89}. This indicates that
correlations in disordered potential though renormalize the localization
length, but does not change the form of the distribution in the weak disorder
limit.
The increasing the strength of the disorder modifies the short time behavior of
 ${\cal P}(\tau)$  (Fig.~\ref{fig2}) . However the typical $1/\tau^2$ tail of
the distribution remains unchanged. This result is again very similar to one
obtained for a one-dimensional Anderson model \cite{OKG00}.

\begin{figure}[t]
\epsfig{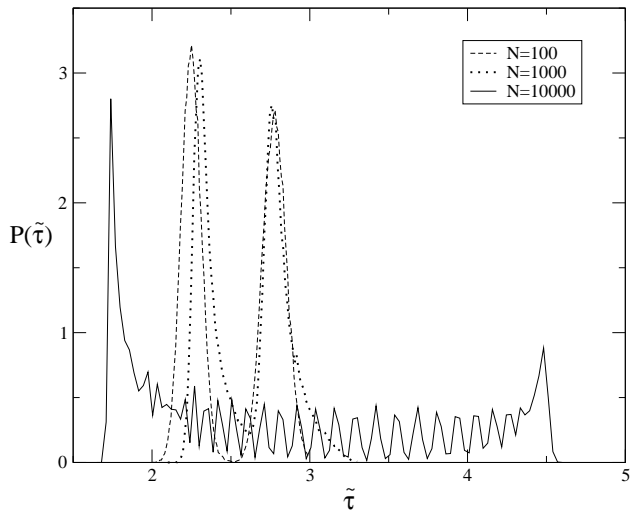}
\caption{\label{fig4} Distribution of the rescaled Wigner delay
times
  $\tilde{\tau}\equiv\tau/N$ for $E=0$, $t=m=0.7$ and  various system sizes $N=100, 1000,
  10000$.}
\end{figure}

Next we concentrate on the critical regime. The most surprising
behavior of ${\cal P}(\tau)$ is found for $t=m=1$  and $E=0$. As
it follows from Fig.~\ref{fig3} the distribution of the rescaled
Wigner delay times $\tilde{\tau}\equiv \tau/N$ goes to the
$\delta$-function in the thermodynamic limit. This means that with
the probability one the particle propagates through the sample
ballistically with a fixed velocity. The ballistic propagation  is
compatible with the delocalization taking place at this point. The
deterministic character of the distribution is however not trivial
and can not be deduced from the delocalization only. This becomes
very clear when we choose $t=m<1$. The distribution of the
rescaled times (Fig.~\ref{fig4}) is bounded (absence of long
tails), but not deterministic anymore. In contrast it has a
complicated oscillating structure in the middle and high peaks at
the edges.  The difference in the behavior of the distributions of the
Wigner delay times for $t=m=1$ and $t=m<1$ is another
manifestation of the difference between the critical indices and DOS found
in the previous  section.

\section{Conclusion} We  have formulated and investigated the
generalization of random dimer model to the case of ladder chain.
As in one-channel case we found delocalization point at  $E_c=0$
with two type of critical behavior when $\epsilon_a-\epsilon_b<2$
and $\epsilon_a-\epsilon_b=2$. The comparison of Landauer and
Lyapunov exponents shows that they have the same critical behavior
in full accordance with the articles \cite{ATAF} and \cite{SSS}.

We studied also the distribution of the Wigner delay times for this model.
Our results shows that its behavior ranges from $1/\tau^2$ decay typical for
the localized system to the deterministic one at the critical point for
$\epsilon_a-\epsilon_b=2$.

Our analytical and numerical results can be relevant for photoluminescence
experiments on electronic properties of GaAs-AlGaAs  superlattices \cite{SS21}
and for experiment on a microwave waveguide with inserted correlated scatterers
\cite{Stoeck}.

At the end we would like to make a following remark. The dimer
model with the correlated disorder defined here does not belong to
the classification of fully disordered systems on the basis of
symmetric spaces \cite{Zirn, Cas}. This is because the geometry of
integration space over random variables, due to the restriction
imposed by correlations, does not coincide with geometry of
symmetric spaces. The classification of universality classes of
the correlated disordered systems is an open and interesting problem.


{\large{Acknowledgments}}: \pagestyle{plain} \makeatletter We wish
to thank V. E. Kravstsov, B. N. Narozhny and O. M. Yevtushenko for
valuable discussions and comments.


\begin{thebibliography}{}

\bibitem{And58}   P. W. Anderson, Phys. Rev.  {\bf 109}, 1492 (1958); N. F. Mott
  and W. D. Twose, Adv. Phys. {\bf 10}, 107 (1961); R. E. Borland,
  Proc. R. Soc. Lond. A {\bf 274}, 529 (1963).
\bibitem{Efet}   K. B. Efetov, {\it Supersymmetry in Disorder and Chaos},
                                   (Cambridge University Press 1997).
\bibitem{Dyson}  F. J. Dyson, Phys. Rev. {\bf 92}, 1331, (1958); G. Theodorou
  and M. H. Cohen, Phys. Rev. B {\bf 13}, 4597 (1976)
\bibitem{Mud1}  P. W. Brower, C. Mudry, B. D. Simons, and A. Altland,
  Phys. Rev. Lett. {\bf 81}, 862, (1998).
\bibitem{Mud2}  P. W. Brower, C. Mudry, and A. Furusaki,
  Nucl.Phys. {\bf B 565}, 653, (2000).

\bibitem{Hein} J. Heinrichs Phys. Rev. B {\bf 66}, 155434 (2003).

\bibitem{SS3}  J. C. Flores, J. Phys. Condens. Matter {\bf 1},
               8471(1989).
\bibitem{DWP}   D. H. Dunlap, H.-L. Wu, and P. Phillips,
              Phys. Rev. Lett. {\bf 65}, 88 (1990).
\bibitem{7SS}   P. Phillips and H.-L. Wu, Science {\bf 252}, 1805 (1991).

\bibitem{SS6}  H.-L. Wu and P. Phillips, Phys. Rev. Lett. {\bf 66}, 1366 (1991).

\bibitem{SS9}  H.-L. Wu, W. Goff, and P. Phillips, Phys.
                Rev. B {\bf 45}, 1623 (1992).
\bibitem{DGK1}  P. K. Datta, D. Giri, and K. Kundu, Phys. Rev. B {\bf 47}, 10727 (1993);
P. K. Datta, D. Giri, and K. Kundu, Phys. Rev. B {\bf 48}, 16347 (1993);
S. N. Evangelou and A. Z. Wang, Phys. Rev. B {\bf 47}, 13126 (1993).
\bibitem{IK}  F. M. Izrailev and A. A. Krokhin, Phys. Rev. Lett. B {\bf 82}, 4062 (1999),
       L. Tessieri and F. M. Izrailev, Physica E {\bf 9}, 405 (2001).
\bibitem{ML}   F.A.B.F. de Moura and M. L. Lyra, Phys. Rev. Lett. {\bf 81}, 3735 (1998).

\bibitem{VP} I. Varga, and J. Pipek, J.Phys.: Condens. Matter {\bf 10}, 305 (1998).

\bibitem{P-D}   D. Sedrakyan and A. Sedrakyan, Phys. Rev. B{\bf 60},10114
                (1999).
\bibitem{SS1}    T. Hakobyan, D. Sedrakyan, A. Sedrakyan, I. Gomez, and F. Dominguez-Adame,
Phys. Rev. B {\bf 61}, 11432 (2000)
\bibitem{SS}    F. Dominguez-Adame, I. Gomez, A. Avakyan,
                D. Sedrakyan, and A. Sedrakyan, Phys. Stat. Sol.(b)
                {\bf 221}, 633 (2000).
\bibitem{TS}    T. Sedrakyan, Phys. Rev. B {\bf 69}, 085109 (2004).
\bibitem{DZL}  L. I. Deych, D.Zaslavsky, and A. A. Lisyansky, Phys. Rev. Lett. {\bf 81}, 5390 (1998).
\bibitem{DL}  L. I. Deych, M. V. Erementchouk and A. A. Lisyansky, Phys. Rev. B {\bf 67}, 024205 (2003).
\bibitem{IKT95} F. M. Izrailev, T. Kottos, and G. P. Tsironis, Phys. Rev. B {\bf
  52}, 3274 (1995); T. Kottos, G. P. Tsironis, and F. M. Izrailev, J. Phys.:
  Condens. Matter, {\bf 9}, 1777 (1997).
\bibitem{IKT96} F. M. Izrailev, T. Kottos, and G. P. Tsironis, J. Phys.:
  Condens. Matter, {\bf 8}, 2823  (1996).
\bibitem{SS21}  V. Bellani, E. Diez, R. Hey, L. Toni, L. Tarricone,
                G. B. Parravincini, F. Dominguez-Adame, and R. Gomes-Alcala,
                Phys. Rev. Lett. {\bf 82}, 2159 (1999).
\bibitem{Stoeck} U. Kuhl, F. M. Izrailev, A. A. Krokhin, and H. -J. Stoeckmann, Appl. Phys. Lett. {\bf 77} 633, (2000).
\bibitem{MI}  F. M. Izrailev and N. M. Makarov, Phys. Rev. B {\bf 67}, 113402 (2003),
    F. M. Izrailev and N. M. Makarov,  Phys. Stat. Sol.(c) {\bf 0}, 3037 (2003).
\bibitem{Zirn}M.Zirnbauer, J. Math.Phys. {\bf 37}, 4986 (1996);\\
A.Altland and M.Zirnbauer, Phys.Rev.Lett. {\bf 76}, 3420 (1996).
\bibitem{Cas}M.Caselle, Phys.Rev.Lett. {\bf 74}, 2776 (1995); cond-mat/9610017.
\bibitem{CPV}  A. Crisanti, G. Paladin, and A. Vulpiani, {\it Products of random matrices},
(Springer-Verlag, 1993).
\bibitem{MacK1} A. MacKinnon, and B. Kramer, Phys. Rev. Lett {\bf 47}, 1546 (1981)
\bibitem{O}    V. I. Oseledec, Trans. Moscow Math. Soc. {\bf 19},197
               (1968).
\bibitem{ATAF} P. Anderson, D. Thouless, E. Abrahams, and D. Fisher,
               Phys. Rev. B {\bf 22},3519 (1980),\\
               P. Anderson, Phys. Rev. B {\bf 22},4828 (1981).
\bibitem{SSS}  R. Schrader, H. Schulz-Baldes and A. Sedrakyan,
               math-phys/0405019, to appear in Annales H. Poincare
\bibitem{KW} M. Kappus, and F. Wegner, Z. Phys. B {\bf 45},15 (1981).
\bibitem{FS97} Y. V. Fyodorov, H.-J. Sommers, J. Math. Phys. {\bf 38} 1918
  (1997), V. A. Gopar, P. A. Mello, M. B\"uttiker, Phys. Rev. Lett. {\bf 77},
  3005 (1996); P. W. Brouwer, K. M. Frahm, C. W. J. Beenakker, ibid. {\bf 78},
  4737 (1997); H.-J. Sommers, D. V. Savin, V. V. Sokolov., ibid. {\bf 87},
  094101
\bibitem{OKG00} A. Ossipov, T. Kottos, and T. Geisel, Phys. Rev. B {\bf 61},
11411 (2000)
\bibitem{JVK89} A. M. Jayannavar, G. V. Vijayagovindan, N. Kumar, Z. Phys. B
{\bf 75}, 77 (1989); C. Texier and A. Comtet, Phys. Rev. Lett., {\bf 82},
4220 (1999); S. A. Ramakrishna and N. Kumar, Phys. Rev. B {\bf 61},
3163 (2000)
\bibitem{F03} Y. V. Fyodorov, JETP Lett. {\bf 78}, 250 (2003)
\bibitem{OKG03} A. Ossipov, T. Kottos, and T. Geisel, Europhys. Lett. {\bf 62},
  719 (2003)
\bibitem{SOKG00} F. Steinbach, A. Ossipov, T. Kottos, and T. Geisel,
  Phys. Rev. Lett., {\bf 85}, 4426 (2000); T. Kottos and M. Weiss, ibid. {\bf
  89}, 056401 (2002); J. A. Mendez-Bermudez and T. Kottos, cond-mat/0401098
  (2004)
\end{thebibliography}
\end{document}